# Societal Complexity and Physical Power

J.L.V. Lewandowski[1]

Institute For Collapse Studies

Rue Jean-Jacques Rousseau

Paris, Cedex 75001, FRANCE

## Abstract

As the current thermo-industrial civilization expands, its technological and societal complexities increase. We suggest that physical power, economic activity and societal complexity are linked. A simple, intuitive model based on Systems Dynamics is used as an illustration.

This paper outlines a mathematical model for Tainter's complexity curve (first section). To the author's knowledge, no such mathematical model has been described before. In the second section, the growth in societal complexity is linked to the (global) economic activity which, itself, is sustained by energy flows (power). The implications of the finiteness of *cheap* energy resources are discussed.

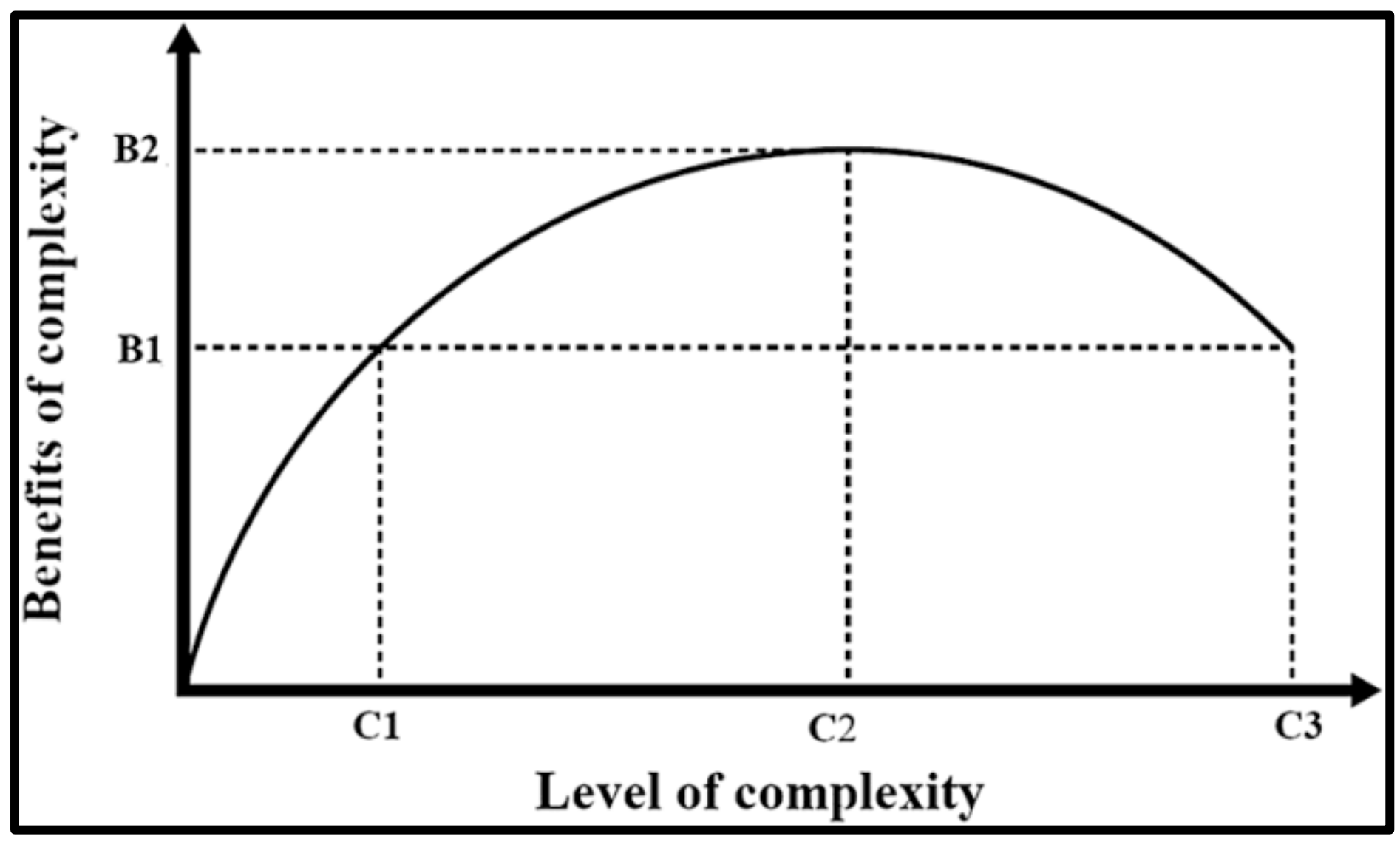

Figure 1. Illustration of Tainter's complexity curve [adapted from Tainter, J. A. (1988). *The Collapse of Complex Societies*. Cambridge University Press].

Mathematical model for Tainter's Complexity Curve

Figure 1 shows the benefits of complexity as a function of the level of complexity, adapted from Tainter's original curve in his book *The Collapse of Complex Societies*. For hyper-complex societies, the cost of complexity rises while its benefits decline. Based on this perspective, we prefer to model the benefit-to-cost ratio as a function of the level of complexity, instead of the benefits (which are ultimately reduced as diminishing returns set it). In the early stages of civilization, the benefits of complexity tend to have a compounding effect (this is not unlike technological growth which acts as a catalyst). As time proceeds, the *growth* in complexity must slow and ultimately reach zero. When the complexity growth rate reaches zero, benefits from complexity have reached their zenith. As a tentative model, we suggest that the *benefits* of complexity, denoted $B$, as a function of the level of complexity, denoted $K$, can be described in the term of a logistic equation as follows:

$$\frac{dB}{dK} = \omega B\left(1 - \frac{B}{B_\infty}\right) \quad (1)$$

[1] InstituteCollapseStudies@gmail.com

Here $\omega$ is a nominal 'growth rate' and $B_\infty$ is the maximum level of benefits that can be achieved. Turning our attention to the *cost* of complexity, we note that in the initial stages of a 'primitive' society or civilization, the cost might vary nearly linearly with complexity. As the internal interconnectedness of society increases, the cost of complexity, denoted $C$, increases faster than linearly; we suggest that it increases according to a power law of the form

$$C(K) \propto K^\beta$$

where $\beta > 1$. For convenience, we rewrite the above equation as

$$C(K) = C_0\left(\frac{K}{K_0}\right)^\beta \quad (2)$$

where $C_0$ and $K_0$ are constants. To understand why the cost of complexity rises at a faster rate than complexity does, we can conceptually conceive a society as a network (an ensemble of nodes and connections). As society expands, both the number of nodes (e.g. human beings and their power-driven artefacts) and the connectivity between them (e.g. faster and cheaper communication) increase: as a result, the number of connections, a rough measure of complexity, rises tremendously. This is illustrated in Figure 2.

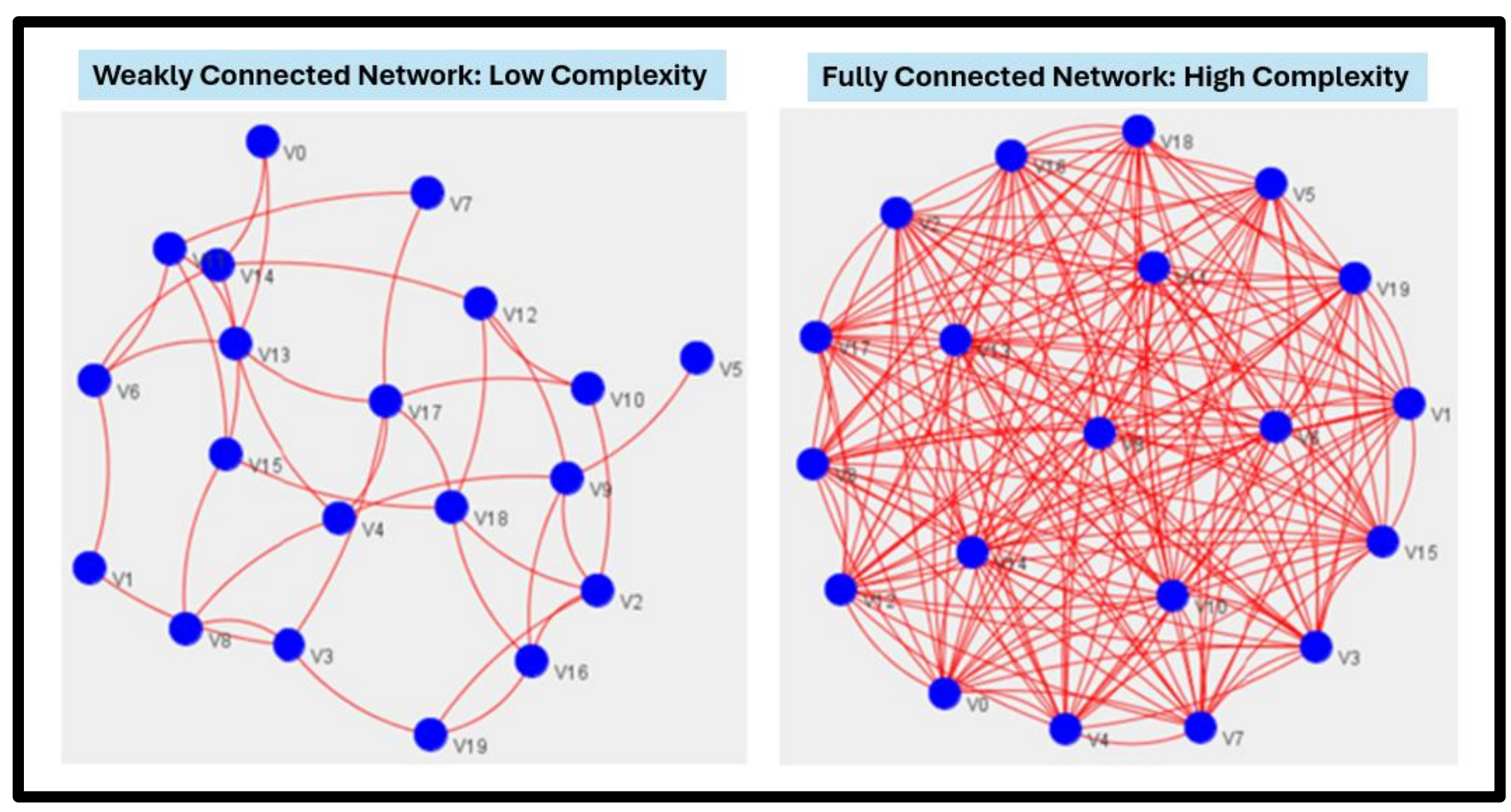

Figure 2: illustration of cost of complexity through the analogy of the connectedness of a network. In the initial stages, a 'primitive' society possesses a weakly connected network of organizations, rules, etc. (left panel). As society grows, interconnectedness increases and complexity cost increases as well (right panel).

As mentioned above, we choose the benefits-to-cost ratio as a proxy for Tainter's 'benefits of complexity' model. The benefits-to-cost ratio is denoted $\theta(K) \equiv \frac{B(k)}{C(k)}$. Equation $(1)$ can be solved exactly and as a result

$$\theta(K) = \theta_0 \frac{\left(\frac{K}{K_0}\right)^{-\beta}}{a + (1-a)\exp\left(-\omega(K - K_0)\right)} \tag{3}$$

where $\theta_0 = \frac{B_0}{C_0} = B(K_0)/C_0$, $a = \frac{B_0}{B_\infty}$ are constant parameters. Figure 3 shows a graphical representation of Equation $(3)$. In the early stages of civilization development, the benefits rise exponentially (due to compounding effects, as noted above) while costs increase modestly. In agreement with Tainter's qualitive picture, as the civilization grows, its complexity increases: problems need more solutions which cause further problems... At a certain stage, benefits from complexity accrue more slowly while costs rise sharply; as a result, the benefit-to-cost ratio of complexity initiates its slow, terminal decline. We must, however, treat the decline of the benefit-to-cost ratio with caution: it is possible for civilizations to undergo systematic simplification in stages through abrupt self-organized reconfigurations. Usually the actual pathways towards 'simplification' are known *a posteriori*.

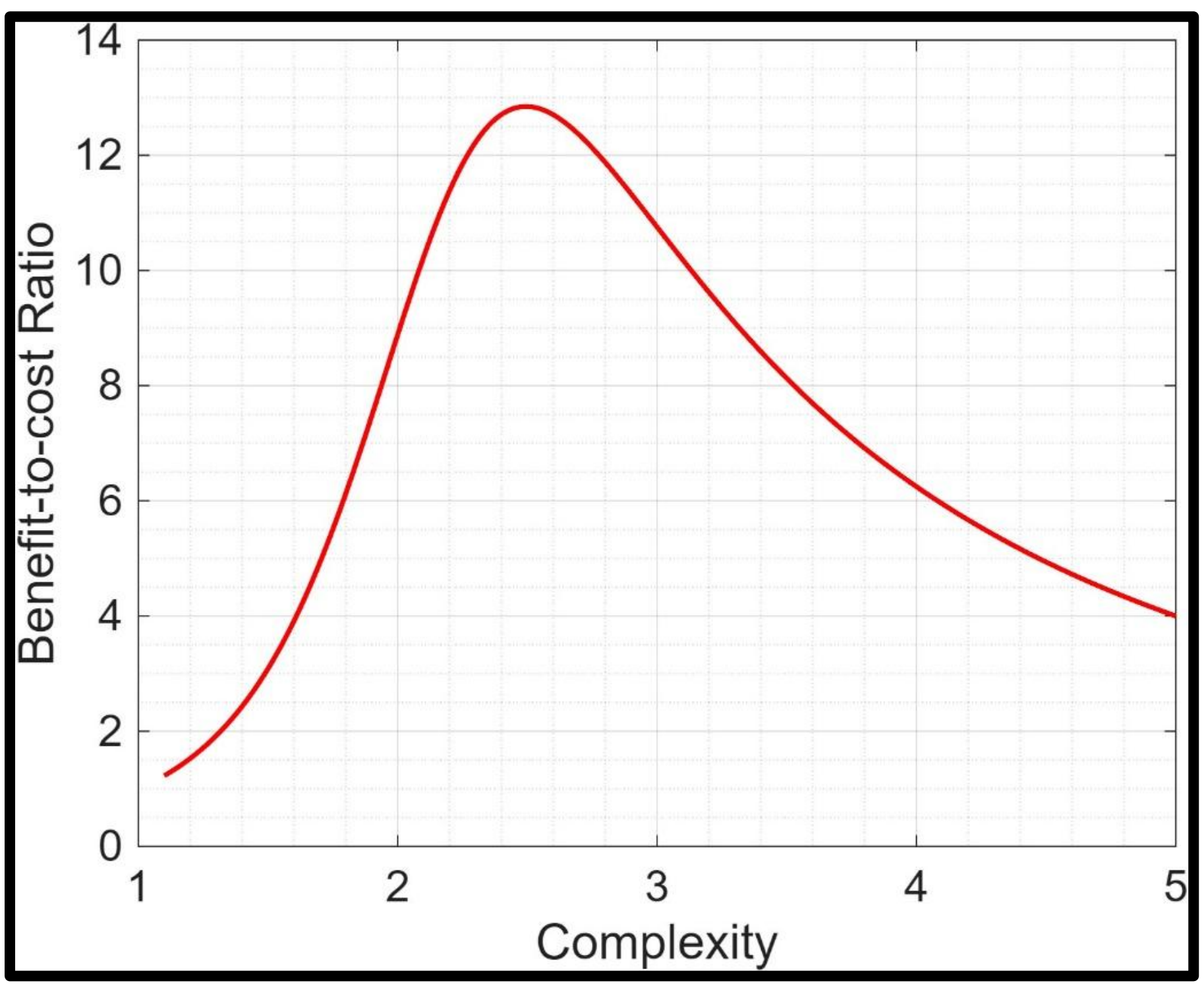

Figure 3. Illustration of the benefit-to-cost ratio of complexity [Equation $(3)$] with parameters $\beta = 2$, $\omega = 4/K_0$ and $a = 0.01$.

Connecting Complexity to Power

Tainter's model is instructive, but it is not connected to socio-economic metrics such total economic activity or physical power. Figure 4 depicts how physical power drives economic activity which, in turn, increases complexity. Green (red) arrows correspond to positive (negative) feedbacks, respectively. The feedback loop between complexity and economic activity is positive in the early stages of a civilization but, ultimately, diminishing returns set it and the feedback loop turns negative (this is shown as a black arrow in Figure 4). Energy stocks (in particular, fossil fuels), being finite, deplete over time; the quality of the remaining energy stocks declines as well. The decline in the *quality* of energy is not accounted in the present model [1].

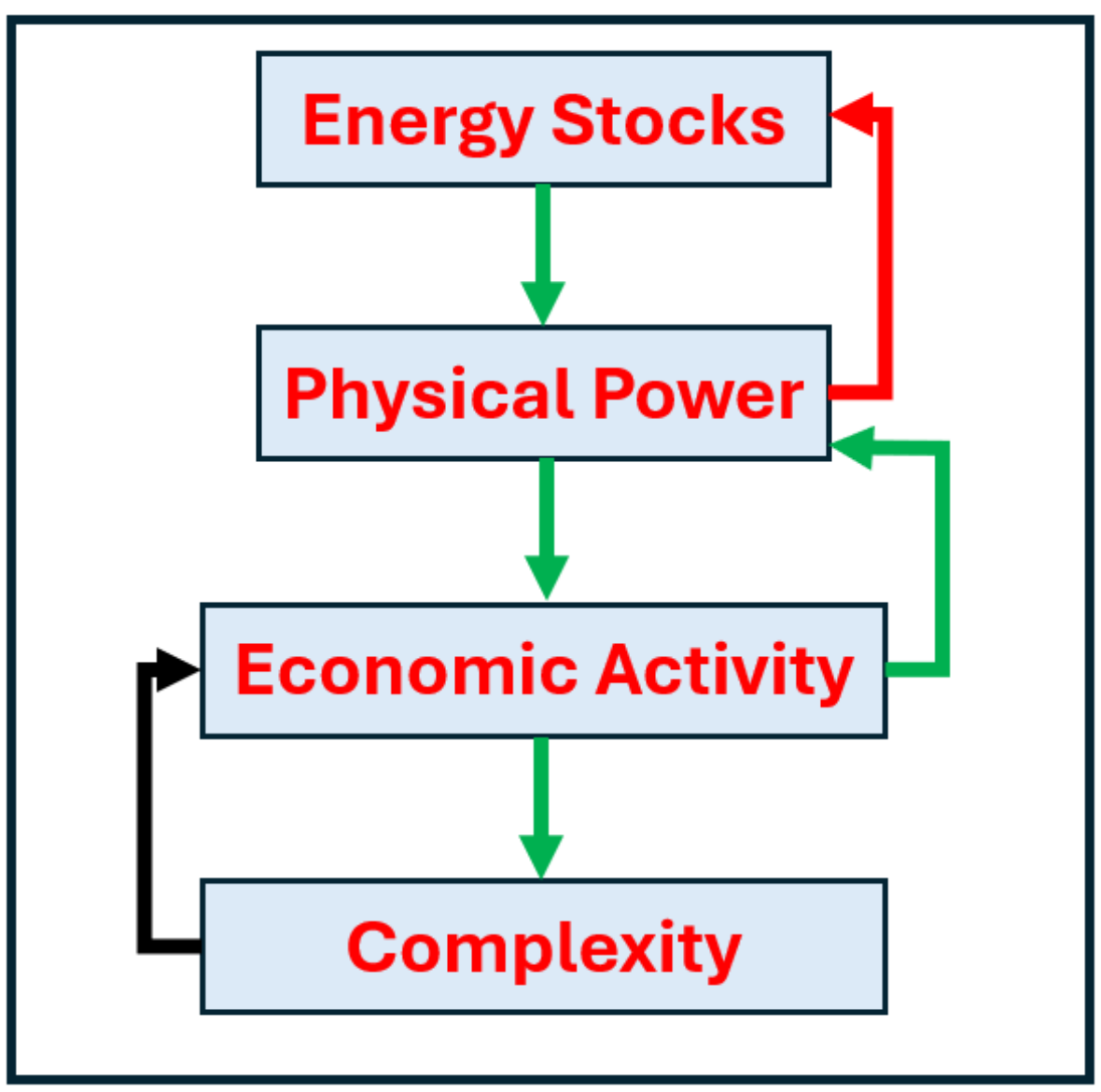

Figure 4: Feedback loops between energy stocks, physical power, economic activity and complexity. Green arrows highlight positive (amplifying) feedback loops whereas red arrows denote negative (stabilizing) feedback loops. The feedback loop between complexity and economic activity (shown as a black arrow) can be positive or negative, depending on the stage of civilizational/societal development.

The strong correlation between economic activity and power (energy per unit of time) is well-established. The sources of power (electricity, natural gas, oil, coal, and renewable sources like biomass, solar, and wind) enable, for example, heating, cooling, transportation, lighting, chemical manufacturing, steel production, aluminum refining and cement manufacturing. Instead of using monetary metrics, it is preferable to quantify economic activity by the energy involved in all (economic) transformations *per unit of time*. Recall that energy per unit of time is (physical) power. The total economic activity is denoted by $S(t)$ which, as noted above, has the units of power. We are adopting an exponential model

$$\frac{dS}{dt} = \gamma(x)S \tag{4}$$

where the growth rate depends on a variable $x$ to be introduced shortly. The choice of an exponential, as in Equation $(4)$, is justified by the fact that the human population has grown at an exponential rate and that power capita has steadily increased over time (one way justify using a faster-than exponential model, but only in a narrow time window, not over the complete civilizational cycle). The power dissipated through economic activity must originate from an input source which we denote as $P$ (physical power); it is reasonable to assume that the overall growth rate must depend on the ratio $P/S$. The variable $x$ in Equation $(4)$ is defined as

$$x = \frac{P}{S} \tag{5}$$

which is unitless since both the economic activity and physical power have the same units. Intuitively, the greater the value $x$ the faster the economic growth rate (The economic boom of the roaring 1950s and 1960s was literally powered by a commensurate extraction and consumption of fossil fuels). However, even if the available physical power is extremely large, the growth rate cannot exceed an upper bound, which we denote $\gamma_M$; after all, the active 'agents' in the economic superorganism are humans and their socio-politico-economic structures which cannot evolve arbitrarily fast (for example, assuming an abundance of cheap, fossil fuels, an airport can be built in a few years, but not a few days). At the same time, a minimum of (physical) power is necessary to simply *maintain* the economic superorganism: for example, even if the economy does not grow, transportation, heating, manufacturing production, lighting, chemical manufacturing, steel production, aluminum refining and cement manufacturing must continue just to *maintain* a given level of economic activity. Therefore, there is a critical value, denoted $x_c$, for which the growth rate vanishes. Therefore the boundary conditions are

$$\gamma(x_c) = 0 \qquad\qquad \gamma(\infty) = \gamma_M \tag{6}$$

The exact form of the growth rate is not known *a priori*. Here we propose that, given an increment $\Delta x$, the corresponding increment for the growth rate is

$$\Delta\gamma = -\mu(\gamma - \gamma_M)\Delta x \tag{7}$$

with positive constant $\mu$. Equation $(7)$ suggests that, as the available power becomes larger (for a given level of economic activity), the increment in growth rate declines and then becomes zero; this is consistent with the above remark that the economic superorganism cannot grow at an arbitrarily fast pace. Integrating Equation $(7)$, and using the boundary conditions $(6)$, we find

$$\gamma(x) = \gamma_M[1 - \exp(-\mu(x - x_c)] \tag{8}$$

Equations $(4)$,$(8)$ constitute the self-reinforcing feedback loop between physical power and economic activity shown in Figure 4. We now turn our attention to the cumulative raw energy available for global economic activity. The main energy sources are fossil fuels (oil, natural gas, coal). Although the wind and solar energy sources are being added at a rapid pace over the past decade, they represent a small fraction of the total global primary energy consumption (a few percent of the total annual power budget). Further, the manufacturing of wind turbines and solar panels require significant amounts of fossil fuels. As a first approximation, we restrict our attention to fossil fuels. Taken in the aggregate, the total quantity of *economically viable* energy sources is finite and denoted $E_T$. The *rate* of energy extraction rises fast in the early stages of civilization expansion, reaches a peak and declines thereafter. The simplest model describing the remaining energy resources is

$$\frac{dE}{dt} = \Omega E\left(1 - \frac{E}{E_T}\right) \tag{9}$$

where $\Omega$ is a nominal growth rate. Equation (9) shows that the available power initially increases at an exponential rate; this is followed by a peak and a decline. The last, and possibly the weakest, key element in our simplistic model is the link between economic activity and complexity. We postulate that complexity and economic activity are related through a power law $K \propto S^{\alpha}$ where the exponent $\alpha$ is positive (this is not to be confused with the parameter involved in the formulation of the growth rate above). The ubiquity of power laws in this context has been hinted in Reference [2] and found its quantitative justification in the work of Hidalgo and Haussman [3]. The key is, of course, the value of the exponent $\alpha$. As the level of economic of activity doubles (say), societal hierarchies do not double in complexity: this can be seen as the "economies' of scale" applied to societal organization. This remark suggests $\alpha < 1$. Therefore, the connection between economic activity and complexity is assumed to be of the form

$$K = K_0\left(\frac{S}{S_0}\right)^{\alpha} \tag{10}$$

where $K_0$ and $S_0$are constants parameters. Before presenting some numerical experiments based on the coupled model (Figure 4), let us discuss the implications of the finiteness of energy resources. Equation (9) suggests the available power initially rises exponentially, reaches a maximum and declines thereafter. Technology can somewhat delay the time of peak power but not eliminate it. The point is that economically viable energy resources are finite (Figure 5).

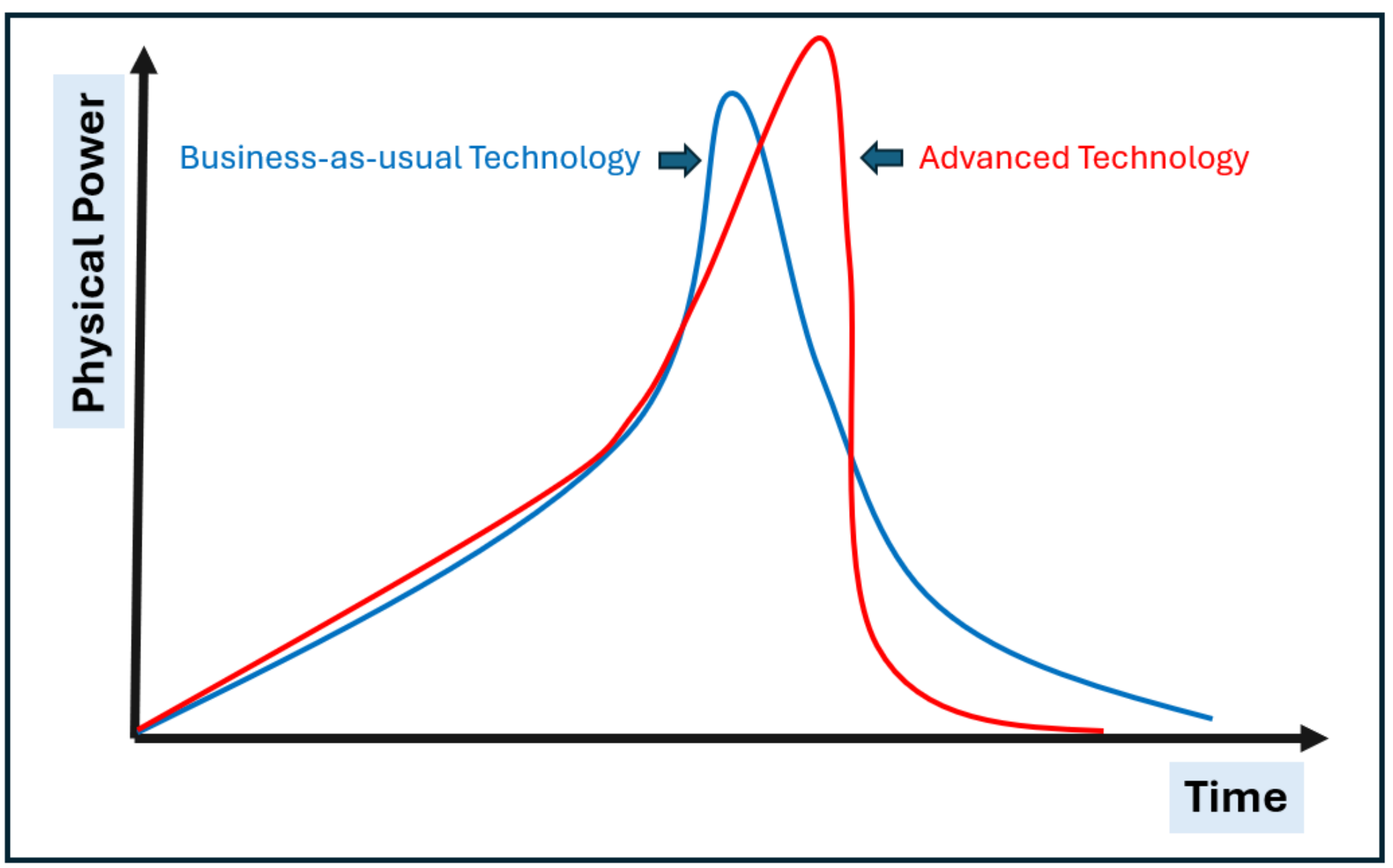

Figure 5: Improving technology can delay the time of peak power but cannot eliminate it. Since the economically viable energy resources are finite, the area under the curve (roughly equal to $E_T$) for both scenarios must be nearly equal (see, for example, Figure 4 in Reference [1]).

As the civilization reaches peak power, it has also reached an enormous size and its complexity is high. As available power inevitably declines, the economic growth rate declines while complexity remains high or even increases slightly. As available power declines further, economic contraction ensues and the society is forced to simplify (de-complexify): this process might be gradual at first, but then accelerates, possibly in a chaotic manner. We now present some numerical experiments based on the coupled model of Figure 4. Figure 6 shows the time evolution of the level of economic activity and its growth rate, power and complexity; we use normalized variables as the shapes of the individual curves that matter. As physical power cannot grow indefinitely, the level of economic activity reaches a peak and decline thereafter. We did not include any delay in our model; doing so would have allowed for societal complexity to grow even past the point of maximum economic activity. Is it worth noting that, in early phase of economic expansion and growth, the (instantaneous) economic growth rate increases over time: the economic activity grows faster than exponentially. This is reminiscent of the 'Trente Glorieuses' [4], a period of massive (and unsurpassed) explosion of economic prosperity in Europe and elsewhere. We must caution that the dynamics of societal complexity once the civilization has entered its decline phase might be very different from the one shown in Figure 6. It is conceivable that the drop in societal complexity might be a slow decline staggered with a series of plateaus, followed by a near steady state, at which point the civilization has self-organized itself into a simpler, more sustainable one.

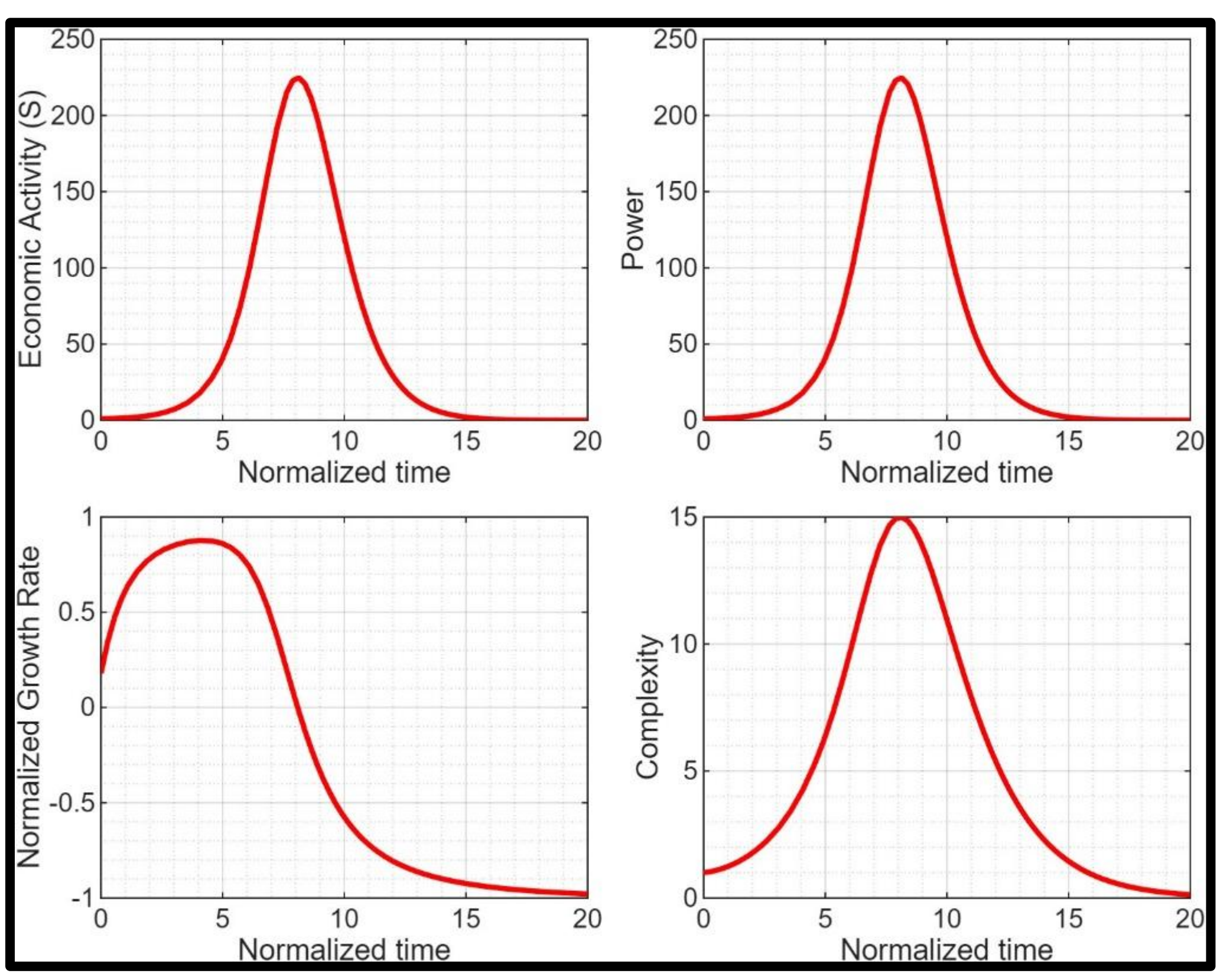

Figure 6. Economic activity, power, economic growth rate and complexity as a function of time for the coupled model. The parameters are $\mu = 1$, $x_c = 0.8$, $\frac{\Omega}{\gamma_M} = 1$. Time is normalized to $\tau = 1/\gamma_M$, the typical time scale of economic growth.